%% file: main.tex
\documentclass[sigconf]{acmart}

\usepackage{url}
\usepackage{amsmath}

\begin{document}

\title{Learning to Evade Static PE Machine Learning Malware Models via Reinforcement Learning}

	\author{Hyrum S.~Anderson}
	\affiliation{ \institution{ Endgame, Inc.} }
	\email{hyrum@endgame.com}

	\author{Anant Kharkar}
	\affiliation{\institution{University of Virginia}}
	\email{agk7uc@virginia.edu}
	
	\author{Bobby Filar}
	\affiliation{ \institution{ Endgame, Inc.} }	\email{bfilar@endgame.com}

	\author{David Evans}
	\affiliation{\institution{University of Virginia}}
	\email{evans@virginia.edu}

	\author{Phil Roth}
    \affiliation{ \institution{ Endgame, Inc.} }	\email{proth@endgame.com}

\begin{abstract}
\input{abstract}
\end{abstract}

	
	


\maketitle

\section{Introduction}
\label{sec:intro}
\input{introduction}

\section{Background}
\label{sec:background}
\input{background}

\section{Method}
\label{sec:method}
\input{method}

\section{Experimental Setup}
\label{sec:expsetup}
\input{expsetup}

\section{Results}
\label{sec:results}
\input{results}

\section{Discussion}
\label{sec:conclusion}
\input{conclusion}

\bibliographystyle{abbrv}
\small
\bibliography{ref}
\end{document}

%% file: abstract.tex
Machine learning is a popular approach to signatureless malware detection because it can generalize to never-before-seen malware families and polymorphic strains.  This has resulted in its practical use for either primary detection engines or for supplementary heuristic detection by anti-malware vendors.  Recent work in adversarial machine learning has shown that deep learning models are susceptible to gradient-based attacks, whereas non-differentiable models that report a score can be attacked by genetic algorithms that aim to systematically reduce the score.
We propose a more general framework based on reinforcement learning (RL) for attacking static portable executable (PE) anti-malware engines.  The general framework does not require a differentiable model nor does it require the engine to produce a score.  Instead, an RL agent is equipped with a set of functionality-preserving operations that it may perform on the PE file.  Through a series of games played against the anti-malware engine, it learns which sequences of operations are likely to result in evading the detector for any given malware sample. This enables completely black-box attacks against static PE anti-malware, and produces functional evasive malware samples as a direct result.  We show in experiments that our method can attack a gradient-boosted machine learning model with evasion rates that are substantial and appear to be strongly dependent on the dataset.  We demonstrate that attacks against this model appear to also evade components of publicly hosted antivirus engines.  Adversarial training results are also presented: by retraining the model on evasive ransomware samples, a subsequent attack is 33\% less effective. Importantly, we release an OpenAI gym to allow researchers to study evasion rates against their own machine learning models, malware samples, and their own RL agents.  We also outline practical limitations with this approach that we hope will beneficial to future research.

%% file: introduction.tex
Machine learning offers an attractive tool for antivirus vendors for both primary detection engines and supplementary detection heuristics. 
Supervised learning models automatically summarize complex relationships among features in the training dataset that are discriminating between malicious and benign labels. 
Furthermore, properly regularized machine learning models generalize to new samples whose features and labels follow the same distribution as the training data set. 

However, motivated and sophisticated adversaries intentionally seek to evade antivirus engines, be they signature-based or otherwise. In the context of a machine learning model, an attacker's aim is to discover and exploit a set of features that the model deems discriminating, but may not be a causal indicator of benign behavior.  The attacker attempts to camouflage the malware in feature space by inducing a feature representation that is highly correlated with, but not necessarily indicative of benign behavior.

In this paper, we present a reinforcement learning (RL) approach to learn to bypass machine learning antivirus models based on static features.  Static detection of malware is an important protection layer in security suites because it allows malicious files to be detected prior to execution. Our intent in attacking machine learning malware models is two-fold: to provide an automated framework to summarize the weaknesses of an anti-malware engine, and to produce functioning evasive malware samples that can be used to augment a machine learning model in adversarial training~\cite{goodfellow2014explaining}.  We focus on static Windows PE malware evasion that presents some unique challenges for a realistic implementation.  However, much of our contributions could be applied to other static machine learning malware detection engines, including PDFs, Mach-O binaries, ELF binaries, etc.
Our implementation is released as an open source Open-AI gym~\cite{brockman2016openai} to enable other  researchers to use, adapt, and improve upon this generic approach.

Several recent works have proposed methods for attacking malware machine learning models in information security~\cite{grosse2016adversarial,anderson2016deepdga,xu2016automatically,hu2017generating,papernot2015distillation}. We present relevant background information, including a review of related work, in Section~\ref{sec:background}.  Unique to our RL approach include (1) the ability to generate functioning Windows PE malware as part of the attack process, which is not possible using gradient-based approaches except under strict and sometimes unrealistic assumptions, (2) the ability to attack a black-box model that does not report a score, and (3) creation of an RL model that can be used to create evasive malware variants from samples not used during training.  We previously released a whitepaper \cite{anderson2017evading} that initially demonstrated these results.  This paper provides additional insight into that work, extends results, and points out limitations discovered since the whitepaper release.

\paragraph{Contributions} The main contributions of this paper include:
\begin{itemize}
\item We present (Section~\ref{sec:method}) a generic black-box attack on static PE malware detection. It assumes no knowledge of the malware model's features or its structure, and requires only the ability to retrieve a malicious/benign label (no score required) for an arbitrary input.  Our attack is based on reinforcement learning, and represents a new direction in automatic evasion research. 

\item We demonstrate evading a machine learning model in Section~\ref{sec:expsetup} with results presented in Section~\ref{sec:results}.
\item We test in Section~\ref{sec:results} whether adversarially-crafted malicious samples can be used to harden a machine learning model via adversarial training. In particular, by retraining a machine learning model using evasive ransomware variants, the evasion rate of a new ransomware attack drops from 12\% to 8\%.
\item We release code\footnote{\url{https://github.com/endgameinc/gym-malware}} in the form of an OpenAI Gym \cite{brockman2016openai} for malware manipulation that may be used by both machine learning and security researchers.
\item Importantly, we point out current practical limitations of this framework first presented in  \cite{anderson2017evading}, that include challenges in guaranteeing functionality of evasive variants, and challenges in using evasive variants for adversarial re-training.
\end{itemize}

%% file: background.tex
This section provides background and summarizes related work in machine learning malware detection and evasion. Section~\ref{sec:staticpe} reviews some common considerations for statically classifying PE files as malicious or benign using supervised learning. In Section~\ref{sec:relatedwork}, we review approaches that have been demonstrated in the literature to attack machine learning models in information security.  

\subsection{Static PE Malware Detection}\label{sec:staticpe}

Static malware detection attempts to classify samples as malicious or benign without executing them, in contrast to dynamic malware detection which detects malware based on is runtime behavior including time-dependent sequences of system calls for analysis~\cite{dahl2013large,pascanu2015malware,athiwaratkun2017malware}. Although static detection is well-known to be undecidable in general~\cite{cohen1987computer}, it is an important protection layer in a security suite because when successful, it allows malicious files to be detected prior to execution. 

Machine learning-based static PE malware detectors have been used since at least 2001~\cite{schultz2001data}, and owing largely to the structured file format and backwards-compatibility requirements, many concepts remain surprisingly similar in subsequent works~\cite{kolter2004learning, shafiq2009framework, raman2012selecting, dahl2013large, saxe2015deep}.  Schultz et al.~\cite{schultz2001data} assembled a dataset and generated labels by running through a McAfee virus scanner.  PE files were represented by features that included imported functions, strings and byte sequences.  Various machine learning models were trained and validated on a holdout set. Models included rules induced from RIPPER~\cite{cohen1995fast}, na\"{i}ve Bayes and an ensemble classifier. Kolter et al.~\cite{kolter2004learning} extended this approach by including byte-level N-grams, and employed techniques from natural language processing, including tf-idf weighting of strings.  Shafiq et al.~\cite{shafiq2009framework} proposed using just seven features from the PE header, including DebugSize, user-definable ImageVersion, ResourceSize, and virtual size of the second listed section, motivated by the fact that malware samples typically exhibit those elements.  Saxe and Berlin leveraged novel two dimensional byte entropy histograms that is fed into a multi-layer neural network for classification~\cite{saxe2015deep}.  

Recent advances in end-to-end deep learning have dramatically improved the state of the art especially in object classification, machine translation and speech recognition.  In many of these approaches, raw images, text or speech waveforms are used as input to a machine learning model which infers the most useful feature representation for the task at hand.  However, despite successes in other domains
hand-crafted features apparently still represent the state of the art for malware detection in published literature.  The state of the art may change to end-to-end deep learning in the ensuing months or years, but hand-crafted features derived from parsing the PE file may continue to be relevant indefinitely because of the structured format.

\subsection{Evading Machine Learning Models} \label{sec:relatedwork}

An attacker may successfully evade machine learning models under a variety of conditions, which may include the following.
\begin{enumerate}
\item Concept drift: an attacker exploits the fact that a model has been trained to approximate $p(\mathbf{x},y)$, but that the concept of malicious vs. benign has since drifted to (or has always actually been represented by) a different distribution $\hat{p}(\mathbf{x},y)$.
\item Modeling error: an attacker exploits the fact that a discriminative model approximates the true posterior $p(y|\mathbf{x})$ as $\hat{p}(y|\mathbf{x})$, and discovers an evasive variant $\mathbf{x}^*$ for which $p(y=\mathrm{malicious}|\mathbf{x}^*)$ $>$ $\eta$, but $\hat{p}(y=\mathrm{malicious}|\mathbf{x}^*)$ $<$ $\eta$ for some chosen threshold $\eta$.  
\item Bayes error rate: even for perfect $\hat{p}(\mathbf{x},y) = p(\mathbf{x},y)$, there is lower bound on the error rate that can be achieved by any machine learning model.  This rate is greater than zero whenever the class-conditional likelihoods $p(\mathbf{x}|y=\mathrm{malicious})$ and $p(\mathbf{x}|y=\mathrm{benign})$ overlap.  This means that there is irreducible error in any model, and a subsequent tradeoff between false positive and false negative rate.  Because of this, in machine learning malware classifiers, the threshold $\eta$ is often set conservatively to avoid false positives due to the base rate of benign samples. (A normally low FP rate may still induce an overwhelming number of false positives due to the high extremely large proportion of benignware on a customer system.)  Irreducible error presents an attack surface for evasion by $\mathbf{x}^*$ by exploiting conservative thresholds $\eta$: $p(y=\mathrm{malicious}|\mathbf{x}^*)$ $<$ $\eta$, even though the true likelihoods may dictate that  $p(\mathbf{x}^*|y=\mathrm{malicious}) < p(\mathbf{x}^*|y=\mathrm{benign})$.
\end{enumerate}
\textit{Bayes error rate} may be best reduced by feature engineering that creates more separable class-conditional distributions, but generally cannot be practically reduced to zero.  \textit{Modeling error} can be reduced significantly reduced by laborious dataset curation and labeling,  disciplined model selection, and techniques such as adversarial training to proactively discover and remedy modeling error.  To some extent, adversarial training can also protect against \textit{concept drift}, in the sense that it hardens a machine learning model against worst-case inputs.

Several recent works have addressed attacking machine learning classifiers for malware.  We group them into three categories according to the amount of information available to the attacker:
\begin{enumerate}
\item \textit{Direct gradient-based attacks} in which the model must be fully differentiable and the structure and weights must be known by the attacker.  Poplar models that defined by differential objective functions include logistic regression, support vector machines, and neural networks (deep learning).  Given the model architecture and parameters/weights, the attacker can essentially query the model directly to determine how to most increase the model's loss function (crudely, how best to bypass the model).
\item \textit{White-box attacks against models that report a maliciousness score}. The attacker has no knowledge about the model structure, but has unlimited access to probe the model and may be able to use heuristics that take advantage of the revealed scores to search for evasive variants.  Virtually every machine learning model can produce a score given a query, but deployed models often choose a threshold on that score to determine malicious / benign. 
\item \textit{Binary black-box attacks}. The attacker has no knowledge about the model, but has unlimited access to probe the model.  The model output is a single bit indicating whether the sample is considered benign or malicious.
\end{enumerate}

\subsubsection{Direct gradient-based attacks}
Gradients from a model under attack provide extremely powerful clues to the attacker, which several attacks can exploit to find evasive variants.

The first approach, introduced by Szegedy et al.~\cite{szegedy2013intriguing}, is to perturb the sample $\mathbf{x}$ in the direction that would most increase the loss function $J(\mathbf{x};\mathbf{\theta})$,
\begin{equation}
\mathbf{x}^* = \mathbf{x} + \mathbf{\sigma}\left( \nabla_{\mathbf{x}} J\left(\mathbf{x};\mathbf{\theta}\right)\right) \label{eqn:perturbation}
\end{equation}
The vector function $\mathbf{\sigma}(\cdot)$ is a domain-specific mapping of the input back to the range of acceptable objects. For example, the Fast Gradient Sign method uses $\mathbf{\sigma}(\mathbf{\delta}) = \epsilon \cdot \mathrm{sgn}(\mathbf{\delta})$ so that the perturbation is imperceptible, maximally bounded by a change of $\epsilon$ to any one pixel~\cite{goodfellow2014explaining}.  In general, the aim of these approaches is to affect a change in the sample so that the the label $f(\mathbf{x})$ predicted by the model changes with some ``minor'' perturbation that keeps the new sample $\mathbf{x}+\delta$ in the domain $\mathcal{D}$ of ``valid'' objects (e.g., human-imperceptible or functioning executable file).  For an energy surface $J( f(\mathbf{x}), y; \theta )$ minimized during model training, an adversarial example attempts to discover
\begin{align*}
\arg\min_{\delta} & \quad J\left( f\left(\mathbf{x} + \mathbf{\delta}\right), \bar{y}; \theta \right) \\
& \mathrm{s.t.}\quad \mathbf{x} + \mathbf{\delta} \in \mathcal{D}\\
 & \qquad\; y \ne \bar{y}.
\end{align*}

Grosse et al.\ attack a deep learning Android malware model using gradient perturbation method~\cite{grosse2016adversarial}.  The feature vector $\mathbf{x} \in \{0,1\}^{545333}$ is a large sparse binary vector.  It is perturbed in a way that bounds the total number of changes (using an $\ell_1$ constraint as a convex proxy to $\ell_0$).  Furthermore, $\mathbf{\sigma}(\cdot)$ is implemented as an index set that allows features to be added (never removed), and only if they do not interfere with other features that are already present in the application.  The authors report evasion efficacy from 50\% to 84\%, depending on the model architecture.  Since the work performs the attack only in feature space, malicious files are never generated during this process, however, and the attack assumes there is a way to generate a sample that matches any feature vector.

A second class of gradient-based attacks connects the model under attack to a generator model in a generative adversarial network (GAN)~\cite{goodfellow2014generative}.  Unlike perturbation methods, the generator learns to generate a completely novel sample from a random seed.  Through a series of adversarial rounds, the generator learns to produce samples that appear to be drawn from the benign class-conditional distribution $\hat{p}(\mathbf{x}|y=\mathrm{benign})$ that has been estimated by the model under attack (the \textit{discriminator} in GAN literature).

Like the perturbation method, a mapping function may be required to ensure that generated samples constitute acceptable objects.  For images, this step is typically ignored.  For malware, the mapping onto legitimate PE files that perform the desired malicious function is essential, but has yet to studied in the general case.

Anderson et al.~\cite{anderson2016deepdga} apply this GAN-based attack to a detector of domain generation algorithm (DGA) domains, which attempts to distinguish  human-crafted from algorithmically-generated domain names.  The only constraint on generated domain names is that they contain valid characters, which are trivially encoded into the alphabet of tokens in the neural network.  As such, the mapping into ``legitimate'' characters is automatically encoded.

\subsubsection{Attacks without gradients}
A less powerful attacker has no direct access to the model, which is necessary for the gradient attacks, but can interact with the model as a black-box that produces a score for any query. This score allows an attacker to directly measure  (myopically) the efficacy of any single perturbation.

Xu et al.~\cite{xu2016automatically} leverage the score reported by PDF malware classifiers to derive a fitness function for a genetic programming framework.  To ensure that mutations preserve the desired malicious behavior, an oracle is used to compare the runtime behavior with that of the original seed.  The authors utilize the Cuckoo sandbox as an oracle, and note how it is computationally expensive.  The genetic algorithm found nearly 17K evasive variants from 500 malicious seeds, and achieved 100\% evasion rate against both the PDFrate and Hidost malware classifiers.  

Recently, Huang et al.\cite{huang2018adversarial} demonstrated a white-box attack for Windows PE files, where the feature vector is known to be a binary indicator of static imports.

\subsubsection{Binary black-box attacks}
Finally, in the most generic attack, the anti-malware engine reports only malicious or benign for an input. This is the minimum output an on-line classification service could provide, so represents the most challenging scenario for an attacker.

Dang et al.~\cite{dang2017evading} attack PDF malware classifiers using only their binary malicious/benign decision.  The authors design a scoring function that can assign real-value scores to reflect evasion progress for a given sample.  The scoring function is composed of only binary outcomes obtained from the model under attack and validation from an oracle that ensures that the behavior of the morphed sample remains unchanged.  This scoring mechanism produces a real-valued objective that can be algorithmically exploited via a hill-climbing approach.  The idea
is to measure the number of morphing steps required
to change the detector's label without modifying behavior measured by the oracle.  In experiments, authors effectively morph 100\% of the test samples to evade the classifiers by this approach.  Required for this approach are the model to attack, an oracle to verify that malicious behavior has been preserved, and a mutation engine.  In summary, this approach aims to trade the binary black box problem with a score-producing black box problem.

Recently, Hu and Tan~\cite{hu2017generating} introduced MalGAN to generate PE malware to bypass a black-box static PE malware engine.  The idea is simple: instead of attacking the black box directly, the attacker creates a fully-differentiable surrogate model trained to reproduce outputs observed by probing the target model with corresponding inputs.  Then, the surrogate model is used for gradient computation in a modified GAN to produce evasive malware variants.  The authors report 100\% efficacy in bypassing the target model, and furthermore, demonstrate that retraining with the adversarial examples has limited efficacy.  

The approach takes advantage of the cross-evasion property, previously observed for computer-vision models~\cite{papernot2017practical}.  But the latter work leverages a more straightforward gradient perturbation method to generate samples adversarial to the surrogate model.  These evaded the target models with high probability.  One could reasonably apply the same approach to PE malware evasion, except for the difficulty in ensuring the file format and malware function have been preserved through an appropriate mapping as in Equation \ref{eqn:perturbation}.

A notable limitation of Hu and Tan's approach is that the attacker must know the complete feature space of the target model~\cite{hu2017generating}.  The substitute model is trained and GAN attack is carried out in this feature space.  The authors argue that the feature space may be discovered by the attacker, and use only imported functions in their evaluations.

Rosenberg et al.~recently reported a black-box attack against \textit{dynamic} machine learning models that leverage recursive neural networks as a basis~\cite{rosenberg2017}.  The model under attack is known to use sequences of API calls (as observed in a sandbox, for example) to determine whether a sample is malicious. As with Hu and Tan~\cite{hu2017generating}, a surrogate model is leveraged as the target for the attack.  As part of the attack, the authors do not modify existing API calls, but insert chaff API calls that do not change the functionality of the malware.  They demonstrate that adversarial sequences found against the surrogate model bypass the original model with 100\% evasion rate. 

\subsubsection{How our work differs}
We target the most challenging scenario for an attacker, where there are limitations on the information available to an attacker:

\begin{enumerate}
\item The output from the target classifier is strictly Boolean, declaring only whether a sample is deemed benign or malicious by the classifier.
\item The feature space and structure of the target classifier are completely unknown.
\item There does not exist an external party (such as an oracle) to guarantee that a sample is valid. Thus, there is no mapping function to the space of legitimate PE files.
\end{enumerate}
Furthermore, we show how the attacker may create an evasion model that can be applied to new malware samples not available during training.  This is in contrast to algorithmic approaches that operate on a fixed population of malware samples.

These restrictions present what we believe is the most difficult (but also most realistic) black-box evasion scenario for static Windows PE malware detectors. As a result of the limited information available to the attacker, evasion rates are significantly lower than those of the approaches above.  However, we believe this study most closely follows approaches used by real-world adversaries---systematically probing anti-malware engines in an attempt to capture and summarize blind spots---but accomplishes it at greater scale. Furthermore, since our ultimate goal is to harden machine learning models against adversarial evasion attacks, one may assume that our automated adversary is at least as capable as the most realistic threat for static Windows PE files, and thus use the generated adversarial samples for model hardening.

Importantly, our approach begins with a pool of malicious PE files and attempts binary manipulations that create an evasive variant.  To our knowledge, our approach is the only work to date that produces valid PE malware samples. (The genetic programming approach used by Xu et al.~ \cite{xu2016automatically} produces valid PDF malware samples.)  These samples can be explicitly used by any machine learning model for model hardening, without respect to differences in feature representations used by each model.

%% file: method.tex
We implement our black-box attack using a reinforcement learning approach \cite{sutton1998reinforcement}.  Section~\ref{ref:rl} provides a brief introduction to reinforcement learning. Then, we following sections explain how we apply it to find evasive Windows PE malware.

\subsection{Reinforcement Learning}\label{ref:rl}

A reinforcement learning model consists of an agent and an environment that interact for a sequence of turns (or discrete timesteps).  For each turn $t$, an agent may choose an action $\mathbf{a}_t \in \mathcal{A}$ based on a policy $\pi(\mathbf{a}|\mathbf{s}_t)$ and an observable environmental state vector $\mathbf{s}_t$.  The environment produces a reward $r_t \in \mathbb{R}$ in response to a chosen action as well as new environmental state vector $\mathbf{s}_{t+1}$.  The reward $r_t$ and observed state of the environment $\mathbf{s}_{t+1}$ are fed back to the agent to choose a new action based on policy $\pi(a|\mathbf{s}_{t+1})$.  The agent learns incrementally through a trade-off of exploration and exploitation which actions to produce given the environment's state.  The reward provides the key objective for learning, and notably, may be zero for many turns until a target state is reached through a relatively long series of actions.  The goal of the agent is to derive a policy that maximizes the expected return defined by $V^\pi(\mathbf{s}_t) = \mathbb{E}_{\mathbf{a}_t}[Q^\pi(\mathbf{s}_t, \mathbf{a}_t)|\mathbf{s}_t]$ with $Q^\pi(\mathbf{s}_t, \mathbf{a}_t) = \mathbb{E}_{\mathbf{s}_{t+1:\infty},\mathbf{a}_{t+1:\infty}}[R_t|\mathbf{s}_t, \mathbf{a}_t]$ and $R_t = \sum_{i \ge 0}\gamma^i r_{t+i}$ where $\gamma \in [0,1]$ discounts the amount of reward from future actions.  Early actions that produce no immediate reward but are important to the final outcome are promoted via $V^\pi$ that predicts the long-term reward for a given state.  This function that estimates the expected utility of taking a given action for a given state is called a Q-function.

Deep reinforcement learning was introduced as a framework to play Atari games by reinforcement agents that often exceed human performance \cite{mnih2013playing,mnih2015human}.  Among the key contributions of the deep reinforcement learning framework was its ability, as in deep learning, for the agent to learn a value function in an end-to-end way: it takes raw pixels as input, and outputs predicted rewards for each action.  This learned value function is the basis for so-called deep Q-learning, where the Q-function is learned and refined over hundreds of games. 

More recently, an actor-critic model with experience replay (ACER) has achieved state of the art performance on a series of Atari tasks.  The model's stability and sample efficiency are notable.  In particular, ACER utilizes a deep neural network to learn both a policy model $\pi$ and a Q-function to estimate the state-action value for each state.  Experience replay is used to help the agent efficiently learn these models from relatively few experiences.  For further information about classical actor critic models, and for recent advances in ACER models, the reader is referred to \cite{konda2000actor} and \cite{wang2016sample}, respectively.

\begin{figure}[tb]
\begin{center}
\includegraphics[width=3.3in]{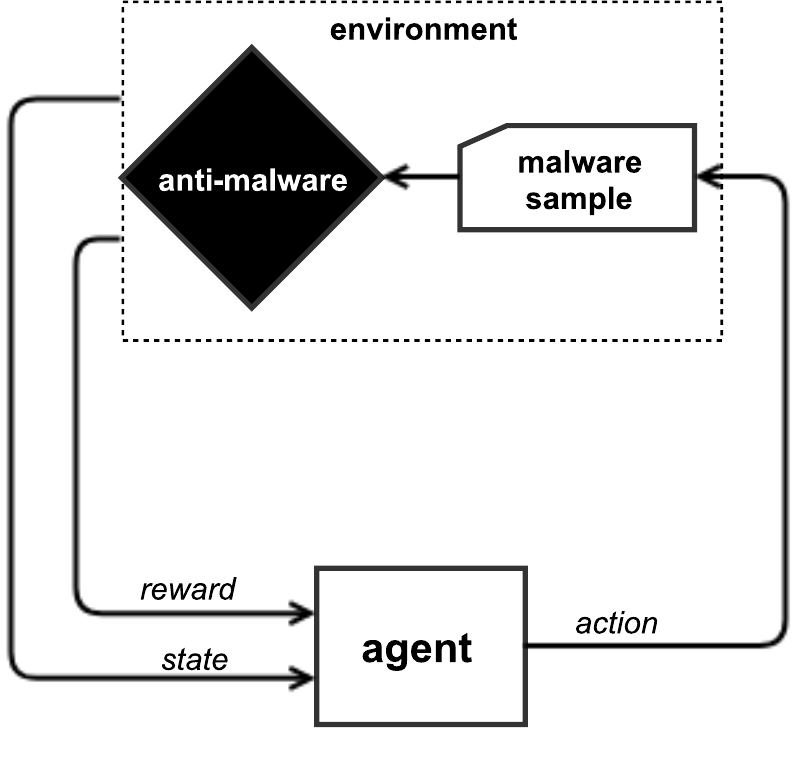} 
\end{center}
\caption{Markov decision process formulation of the malware evasion reinforcement learning problem.}\label{fig:mdp}
\end{figure}

In our experiments, we train an ACER agent to learn a policy for our framework depicted in Figure~\ref{fig:mdp}.  In the Markov decision process shown, the agent gets an estimate of the environment's state $\mathbf{s} \in \mathcal{S}$, represented by a feature vector $\mathbf{s}$ of the malware sample (which need not correspond to any internal representation of the malware by the anti-malware engine).  The Q-function and action policy determine what action to take.  In our framework, the actions space $\mathcal{A}$ consists of a set of modifications to the PE file that (a) do not break the PE file format, and (b) do not alter the intended functionality of the malware sample.  The reward function is measured by the anti-malware engine, which is converted to a reward: 0 if the modified malware sample is judged to be malicious (no evasion), and $R$ if it is deemed to be benign (evasion).  The reward and state are then fed back into the agent.

\subsection{Implementation}

With an aim to engage the broader community, we implement our malware evasion environment as an extensible OpenAI \texttt{gym} \cite{openaigym}, which we release at \texttt{\url{https://github.com/endgameinc/gym-malware}}. 
The \texttt{gym} framework has become popular for training RL agents because it provides a standardized environment to produce benchmarks (like playing Atari games). We adopt some game-playing terminology in some of our description below.  In addition, we release a default ACER agent using \texttt{chainer-rl} \cite{chainer_learningsys2015}.  PE file parsing and manipulation leverages the Library to Instrument Executable Formats \cite{LIEF}.

The environment consists of an initial malware sample (one malware sample per ``game''), and a customizable anti-malware engine (the attack target). Each step or turn provides the following feedback to the agent:
\begin{itemize}
\item A reward value $\in \{0,R\}$, where $0$ denotes the malware sample was detected by the anti-malware engine and $R$ is the reward given for evading the engine.  In our gym and in experiments, we use $R=10$.
\item A feature vector summarizing the state of the environment (malware sample). The feature vector is described in further detail below.
\end{itemize}

Based on this feedback, the agent chooses from a set of mutations \textit{(actions)} that preserve the format and function of the PE file.  We describe our initial implementation of each of these components below. 

\subsection{Environment}
The malware sample exists as raw bytes in the game environment.  However, in order to more concisely represent the current state of the malware sample, the environment emits the state in the form of a feature vector.  In our experiments, we use a 2350-dimensional feature vector comprised of the following general categories of features:
\begin{itemize}
\item PE header metadata
\item Section metadata: section name, size and characteristics
\item Import and Export Table metadata
\item Counts of human readable strings (e.g. file paths, URLs, and registry key names)
\item Byte histogram
\item 2D byte-entropy histogram (as used by Saxe and Berlin~\cite{saxe2015deep})
\end{itemize}

For feature sets which are countably infinite (section names, imported function names, etc.), we use the hashing trick to collapse into them into a vector of fixed size.  The resulting feature vector represents a fairly holistic view of the malware sample, and encapsulates elements used by machine learning malware models in industry and academia~ \cite{kolter2004learning, shafiq2009framework, raman2012selecting, dahl2013large, saxe2015deep}.

These features can be inspected in the accompanying code, and we note that they can be applicable both to training a machine learning malware detector, as well as for our use in summarizing a malware sample that comprises our ``state'' in the Markov decision process.

\subsection{Action Space}
As mentioned above, the file mutations represent the actions or moves available to the agent within the environment. There are a modest number of modifications that can be made to a PE file that do not break the PE file format and do not alter code execution.  Each of these can be inspected in detail in the published code. Some of these include:
\begin{itemize}
\item adding a function to the import address table that is never used (this is the sole manipulation explored by Hu and Tan~\cite{hu2017generating})
\item manipulating existing section names
\item creating new (unused) sections
\item appending bytes to extra space at the end of sections
\item creating a new entry point which immediately jumps to the original entry point
\item removing signer information
\item manipulating debug info
\item packing or unpacking the file
\item modifying (breaking) header checksum
\item appending bytes to the overlay (end of PE file)
\end{itemize}

Note that most of these functions are stochastic in nature.  For example, when renaming a section, a new section name is drawn uniformly from a list of section names found in benign files.  When appending bytes to the end of a section or file, the length and entropy of of the appended bytes can be specified, but for simplicity, are chosen at random by the agent.  Likewise, the compression level used by the packer is chosen at random.

The stochastic nature of the manipulations was chosen for simplicity to reduce an exponentially large number of mutations to a few dozen stochastic actions.  The idea is that the agent may now choose actions that modify broad elements of a PE file which are generally used by static machine learning malware models.  An alternative is to unroll the limited number of actions into hundreds of specific actions (e.g., rename section \texttt{.wah} to \texttt{.blah}, instead of renaming randomly).  However, reinforcement learning with extremely large action spaces is a subject of ongoing research~\cite{dulac2015deep}.

%% file: expsetup.tex
In our experiments, we attack a gradient boosted decision tree model trained on 100,000 malicious and benign samples, and which achieves an area under the receiver operating characteristic score (ROC-AUC) of 0.993.  In our experiments, we set a threshold of 0.9 for the static malware model that approximately corresponds to a 1\% false positive rate at a 90\% true positive rate.  This model is included in the code that we release.  

Although not strictly necessary, for convenience, we train the agent on the features used to represent the state of the environment.  We expect this to produce more generous results than can be expected in practice.  However, as our intent is to release a toolkit for learning malware manipulating agents, this proof of concept suffices for our purposes. In general, we postulate that the agent's learning problem is simpler when each of the following conditions hold: (1) the feature representation used by the model under attack has significant overlap with the features used by the agent to represent the malware state, (2) the agent's actions are fully observable by the state representation---that is, the agent can ``see'' via his feature representation the effect of his actions, and (3) the agent's actions can affect with reasonable granularity any part of the feature vector used by the model under attack. 

Our preliminary experiments involved our basic chainer-rl actor-critic model agent tested in our OpenAI gym. The agent utilizes a Boltzman exploration / exploitation strategy, in which mutations are drawn proportionally to their expected Q-value.  The agent is allowed to perform up to ten mutations before declaring failure (i.e., ten rounds with exactly one mutation per round).  The relatively few number of mutations allows is motivated by a few considerations.  First, long sequences of moves that finally produce a reward can lead to complications in training the reinforcement learning agent (the \textit{credit assignment} problem).  Secondly, we wish to keep the spirit of adversarial perturbations, in which the ``size'' of the perturbation is relatively small.  This is important in our domain because degenerate sequences such as $\mathtt{add\_section},  \mathtt{add\_section},  \mathtt{add\_section}, \ldots, \mathtt{add\_section}$
may indeed bypass the classifier, but also represents a new signature that is fairly indicative of having been produced by our automated method.  In preliminary experiments, we found that allowing for long sequences could promote these degenerate sequences against our particular model.  An alternative approach would be to regularize the entropy of the action space, which we did not deeply explore.

The ``game'' is comprised of several rounds. Each round begins with a known malware sample, which is modified through a series of mutations in the round.  Rounds terminate early if the agent bypasses the malware model prior to the ten allotted mutations (i.e., the agent was bypassed in less than ten mutations). We allow a combined total of 50,000 mutations to train each model.

For the black box attack, rewards of $R=10$ or $R=0$ are provided for evasion or failed-evasion, respectively. For comparison, we use an attack in the same environment using randomized action with no RL agent.

%% file: results.tex
We train four separate agents that differ only in the malware samples used as seeds for evading the machine learning model.  In particular, we train an agent on the following datasets consisting of the following samples:
\begin{itemize}
\item 50K randomly selected malware samples from Virus Share,
\item 2.6K ransomware samples downloaded from VirusTotal,
\item 1.6K Virut samples downloaded from VirusTotal, and
\item 4.1K BrowseFox (adware) samples downloaded from VirusTotal.
\end{itemize}

For each of these datasets, we set aside 200 samples for a holdout validation set, and use the remaining samples for training the RL agent.

During training of the reinforcement learning agent, we save malware samples that result in an evasion.  The number of evasive variants discovered during training with a fixed budget of 50K mutations are summarized for each category in Table \ref{tab:evasion_training}.  Note that during training, if the machine learning model classifies the original malware seed as benign, the sample is skipped and not included in our analysis.

\begin{table}
    \centering
    \begin{tabular}{r||rl}
        \textbf{dataset} & evasions & (\% of budget)\\
        \hline
         \textbf{VirusShare} & 2085 & (4.2\%) \\
         \textbf{ransomware} & 1543 & (3.1\%) \\
         \textbf{Virut} & 619 & (1.2\%)\\
         \textbf{BrowseFox (adware)} & 2444 & (4.9\%) \\        
         \hline
    \end{tabular}
    \caption{Number of evasive variants found during training, where we gave the agent a budget of 50K mutations.}
    \label{tab:evasion_training}
\end{table}

\paragraph{Generalization.} We wish to determine how well the trained reinforcement learning agent can generalize to samples never before seen.  After an agent is trained, we measure performance on 200 holdout samples that have been withheld from both the malware detector training and the agent's training.  In particular, for each of the 200 samples in each dataset, we use the trained agent to manipulate the sample (up to 10 mutations), and check whether any subsequence of mutations results in an invasive variant.  As with the training, if the original malware sample is misclassified as benign, the sample is skipped and not included in our analysis.

We compare evasion rates by the reinforcement learning to a random uniform exploration policy: up to 10 mutations per samples are chosen at random from the set of possible actions. Results are summarized in Table \ref{tab:evasion_results}.

\begin{table}
    \centering
    \begin{tabular}{r||c|c}
        \textbf{dataset} & agent & random\\
        \hline
         \textbf{VirusShare} & 24\% & 23\% \\
         \textbf{ransomware} & 12\% & 9\% \\
         \textbf{Virut} & 10\% & 9\% \\
         \textbf{BrowseFox (adware)} & 19\% & 18\% \\
         \hline
    \end{tabular}
    \caption{Evasion rate on 200 holdout samples.  A samples is included in the calculation only if the classifier correctly identified the original sample as malicious.}
    \label{tab:evasion_results}
\end{table}

\paragraph{Dominant mutations.} We summarize in Table \ref{tab:dominant_action} the dominant mutations selected by both the agent and by the random policy for the holdout samples that successfully evaded the machine learning model.  Note that the both the agent and random policy found the the machine learning malware model struggles with \texttt{upx\_pack}.  However, note that agent generally exploits this blind spot in a fewer number of moves, as denoted by the median number of moves used in a successful evasion in the table.

\begin{table}
\centering
\begin{tabular}{r||c|c}
    \textbf{dataset} & agent & random\\\hline
         \textbf{VirusShare} & \small{\texttt{upx\_pack}} (2) & \small{\texttt{upx\_pack}} (2) \\
         \textbf{ransomware} & \small{\texttt{section\_rename}} (1) & \small{\texttt{imports\_append}} (1.5) \\
         \textbf{Virut} & \small{\texttt{upx\_pack}} (4) & \small{\texttt{upx\_pack}} (5.5) \\
         \textbf{BrowseFox (adware)} & \small{\texttt{upx\_pack}} (1.5) & \small{\texttt{upx\_pack}} (2.5) \\
\end{tabular}
\caption{Dominant successful mutations seleted by the RL agent and by the random policy.  Shown in parentheses is the median number of mutations required for successful evasion.}
\label{tab:dominant_action}
\end{table}

\paragraph{Cross-evasion.} We uploaded samples produced by the reinforcement learning agent to VirusTotal, and found that the median detection ratio dropped when compared to the detection ratio from the original sample.  We emphasize that VirusTotal does not represent the full detection platform for vendors in VirusTotal; nevertheless, the drop in detection rate does demonstrate that by bypassing a relatively simple machine learning model, cross-evasion of commercial products may be possible. For comparison, we also test the cross-evasion efficacy of random mutations on the same data, which surprisingly, is similarly effective at evading a host of anti-virus solitions on VirusTotal.  A summary of cross-evasion efficacy is shown in Table \ref{tab:crossevasion_results}.  Apparently, a host of AV scanners in VirusTotal are susceptible to mutations of malware samples.

\begin{table}
    \centering
    \begin{tabular}{r||c|c||c|c}
        \textbf{dataset} & original & agent & original & random \\
        \hline
         \textbf{VirusShare} & 54/65 & 26/65 & 54/65 & 16/65 \\
         \textbf{ransomware} & 52.5/65 & 16.5/65 & 44.5/65 & 9.5/65 \\
         \textbf{Virut}& 57/65 & 25/65 & 56.5/65 & 20/65 \\
         \textbf{BrowseFox (adware)} & 49/65 & 18/65 & 49/65 & 21/65 \\
         \hline
    \end{tabular}
    \caption{Cross-evasion rates on 200 holdout samples, showing the median number of VirusTotal detections before and after mutation using agent mutations and random mutations.}
    \label{tab:crossevasion_results}
\end{table}

\paragraph{Model hardening.} To test whether a model can be hardened by adversarial training, we took the 1543 samples discovered during training of the \texttt{ransomware} dataset and add them to the 100K training set of the original GBDT model and retrained.  We then used the previously-trained reinforcement learning agent to manipulate the 200 holdout samples (which neither agent or model have seen) and test the evasion efficacy of the retrained model.  In this case, the reinforcement learning agent successfully discovered evasive variants for 8\% (down from 12\%) of the samples, a drop of 33\%.  It is important to reiterate that the attack efficacy is measured only on malware samples that the machine learning model initially labels as malicious.  In this way, we have tried to decouple the fact that the ``hardened'' machine learning malware model (trained on generic malware samples) might merely be more aware of ransomware samples, rather than aware of blind spots in the model.

\paragraph{Shortcomings and open challenges.} These black-box evasion and hardening results are modest compared to other gradient-based and white-box attacks.  Furthermore, upon close inspection of samples, we discovered the following, which we describe for future work:
\begin{enumerate}
\item For a random sampling of ten evasive variants generated from the \texttt{VirusShare} dataset, we discovered that only eight executed properly in a virtual machine.  This was troubling, since our approach was based on the ``functional by construction'' assumption that mutations preserve format and function.  In corresponding with authors of the library we use for binary manipulation \cite{LIEF}, has lead us to believe that while the mutations themselves \textit{should} preserve function and format, they may not respect certain obfuscation tricks or less common uses of the PE file format.  For example, \texttt{putty.exe} uses \texttt{LEA} instructions in the import address table, so that patching with our approach would also require patching assembly code after the table is rebuilt. Furthermore, malware may intentionally exploit lazy parsing by the Windows loader to remain functional (e.g., section sizes), while enjoying some degree of anti-tampering by technically violating the PE standard that parsers rely on.  In this case, our file manipulations that require parsing do not practically work since the parsed file is already invalid prior to employing the manipulation.  One workaround for this issue is to begin with a pool of malware samples that are known to parse correctly.
\item We demonstrated that after adversarial training, a new model was hardened to a subsequent adversarial attack.  However, on close inspection, we observed that the library we use for binary manipulation \cite{LIEF} leaves unique fingerprints in samples it modifies.  For example, although section names are in large part arbitrary, the model we use in this paper relies on section names to build evidence of malicious vs. benign.  In file modified by LIEF, one sees uncommon section names like \texttt{.l1} and \texttt{.l2} that are idiosyncratic.  A danger of adversarial training with these samples is that the model begins to learn the difference between ``modified by LIEF'' and ``not modified by LIEF'', rather than ``malicious'' and ``benign''.
\end{enumerate}
For other researchers pursuing this line of research, we strongly caution an investigation into these matters, which is outside the scope of this research.

%% file: conclusion.tex
We have demonstrated a generic black-box attack based on reinforcement learning against Windows PE machine learning models.  This represents a new direction in automatic evasion research.  Unique to our approach include the following.
\begin{itemize}
    \item The attacker requires no knowledge about the model under attack.
    \item The RL agent can generalize: it can modify new malware samples to bypass the model it was trained against.
    \item We believe our approach to be the first to automatically create novel Windows PE evasive malware variants by modifying binary files.
\end{itemize}

We also demonstrated on a ransomware dataset that a model could be made more robust to a direct reinforcement learning attack by retraining on adversarial examples.  However, we note that these results are dataset and model dependent, and leave a thorough study of model hardening to future work.  Generally, we found that many of the engines in VirusTotal were susceptible to static changes in the malware, whether made by the targeted agent, or just by random mutations.

Evasion rates by this RL approach appear to be exceeded by other approaches for attacking machine learning models in information security, broadly, in which the attacker has more knowledge about the model under attack. We believe, however, that our novel approach best mimics real-world conditions, in which an attacker may have only API access to a model that reports malicious or benign (for example, machine learning models hosted on VirusTotal).  This represents the most difficult scenario for an attacker.  If one assumes that our automated adversary is equally or more capable than attackers in the wild, then this approach represents a genuinely valuable means for generating evasive variants for studying model weaknesses or for adversarial training.

Although this work lays the groundwork for generating evasive variants for black-box attacks, we have also outlined some corner-cases that represent points of failure.  In particular, in some cases, ``functionality-preserving'' mutations actually break the PE file format due to parsing issues related to uncommon practices or intentional obfuscation techniques used by malware.  In addition, care must be taken when retraining on adversarial examples generated by this approach, since these samples may be idiosyncratic in ways unrelated to malicious vs. benign.  (In essence, one could actually begin to poison ones own dataset via adversarial training.)  We leave these important issues for future work, and as a cautionary tales to other researchers.

Lastly, we have provided open source code at \url{https://github.com/endgameinc/gym-malware}.  We believe that there is significant room for improvement in this approach, and encourage researchers to contribute.  Specifically, researchers may find interest in enriching the environment with additional functionality-preserving mutations, improving the reinforcement learning agents used for attack (the purpose of the gym framework), and especially for industry researchers, directing their own agents to attack their own machine learning models.